\documentclass[prc,preprintnumbers,twocolumn]{revtex4}
\usepackage{graphicx}
\usepackage{amssymb}
\usepackage{amsmath}

\begin{document}

\title{Effects of nuclear deformation and neutron transfer
 in capture process, and origin of fusion hindrance at deep sub-barrier energies}
\author{V.V.Sargsyan$^{1,2}$, G.G.Adamian$^{1}$, N.V.Antonenko$^1$, W. Scheid$^3$, and  H.Q.Zhang$^4$
}
\affiliation{$^{1}$Joint Institute for Nuclear Research, 141980 Dubna, Russia\\
$^{2}$International Center for Advanced Studies, Yerevan State University, M. Manougian 1, 0025, Yerevan, Armenia\\
$^{3}$Institut f\"ur Theoretische Physik der Justus--Liebig--Universit\"at, D--35392 Giessen, Germany\\
$^{4}$China Institute of Atomic Energy, Post Office Box 275, Beijing 102413,  China
}
\date{\today}

\begin{abstract}
The roles of  nuclear deformation
and neutron transfer in  sub-barrier capture process are  studied within the quantum diffusion approach.
The change of the deformations of colliding nuclei with neutron exchange can crucially influence the sub-barrier
fusion. The comparison of the calculated capture cross section and
the measured fusion cross section in various reactions
at extreme sub-barrier energies gives us
information about  the fusion  and quasifission.
\end{abstract}


\maketitle

\section{Introduction}
The  nuclear deformation and neutron-transfer process
have been identified as playing a major role in the
magnitude of the sub-barrier capture and fusion cross sections~\cite{Gomes}.
There are a several experimental evidences which confirm the importance
of  nuclear deformation on the capture and fusion.
The influence of nuclear deformation is straightforward.
If the target nucleus is prolate in the ground state,
the Coulomb field on its tips is lower than on its sides, that
then increases the capture or fusion probability at  energies below
the barrier corresponding to the spherical nuclei.
The role of neutron transfer
reactions is less clear. A correlation between the overall transfer
strength and fusion enhancement was firstly noticed  in Ref.~\cite{Henning}.
The importance of neutron transfer with positive $Q$-values
on nuclear fusion (capture) originates from
the fact that neutrons are insensitive to the Coulomb barrier and
therefore they can start being transferred at larger separations
before the projectile is captured by target-nucleus~\cite{Stelson}.
Therefore, it is generally thought
that  the sub-barrier
fusion cross section will increase~\cite{Pengo,Roberts,Stefanini3236s110pd,Acker,Sonzogni}
because of the neutron transfer.
As suggested in Ref.~\cite{Broglia},  the
enhancements in fusion yields may be due to the transfer of a neutron pair with a positive
$Q$-value.
However, as shown  recently in
Ref.~\cite{Scarlassara}, the two-neutron transfer channel with large positive $Q$-value
weakly influences  the fusion
(capture) cross section in the  $^{60}$Ni + $^{100}$Mo  reaction at sub-barrier energies.
So, from the present data an unambiguous signature of the
role of neutron transfer channel  could not be inferred.

The experiments with various medium-light and heavy systems
have shown that the experimental slopes
of the complete fusion excitation function
keep increasing at low sub-barrier energies and may become much larger than the
predictions of standard coupled-channel calculations.
This was identified as the fusion hindrance~\cite{Jiang}.
More experimental and theoretical studies of sub-barrier fusion hindrance
are needed to improve our understanding of its physical reason,
which may be especially important in astrophysical fusion reactions~\cite{Zvezda}.

It is worth remembering that the first evidences of hindrance  for compound nucleus
formation in the reactions with massive nuclei ($Z_1\times Z_2>1600$)
at  energies near the Coulomb barrier
were observed at GSI already long time ago~\cite{GSI}.
The theoretical investigations showed that the probability of complete fusion  depends
on the competition between the complete fusion and quasifission
after the capture stage~\cite{Volkov,nasha,Avaz}. As known, this competition can strongly reduce
the value of the fusion cross section and, respectively,
the value of the evaporation residue cross section
in the reactions producing superheavy nuclei.
Although the quasifission was originally ascribed to the reactions with massive nuclei, it is the general
phenomenon which is related to the binary decay of nuclear system after the capture, but before the compound nucleus
formation which could exist at angular momenta treated. The mass and angular distributions of the
quasifission products depend on the entrance channel and bombarding energy~\cite{Volkov}.
Because the capture cross section is the sum of the fusion and quasifission cross sections,
from the comparison of calculated capture cross sections and measured
fusion cross sections
one can extract the hindrance factor and
show a  role of the quasifission channel
in the reactions with various medium-mass and heavy nuclei at extreme
sub-barrier energies.

In the present paper the quantum diffusion approach~\cite{EPJSub,EPJSub1} is applied to study
the fusion hindrance and
the roles of  nuclear deformation
and neutron transfer in  sub-barrier capture process.
With this approach many heavy-ion capture
reactions at energies above and well below the Coulomb barrier have been
successfully described~\cite{EPJSub,EPJSub1,Conf}.
Since the details of our theoretical treatment were already published in
Refs.~\cite{EPJSub,EPJSub1}, the model will be shortly described in Sec.~II.
The calculated results will be presented in Sec.~III.

\section{Model}
In the quantum diffusion approach
the collisions of  nuclei are treated in terms
of a single collective variable: the relative distance  between
the colliding nuclei. The  nuclear deformation effects
are taken into consideration through the dependence of the nucleus-nucleus potential
on the deformations and orientations of colliding nuclei.
Our approach takes into consideration the fluctuation and dissipation effects in
collisions of heavy ions which model the coupling with various channels.
We have to mention that many quantum-mechanical and non-Markovian effects accompanying
the passage through the potential barrier are taken into consideration in our formalism \cite{EPJSub,our,VAZ}.
The details of  used formalism are presented in our previous  articles~\cite{EPJSub,EPJSub1}.
All parameters of the model are set as in Ref.~\cite{EPJSub}.
All calculated results are obtained with the same set of parameters and are rather insensitive
to the reasonable variation of them~\cite{EPJSub,EPJSub1}.
The heights of the calculated Coulomb barriers $V_b=V(R_b)$
($R_b$ is the position of the Coulomb barrier)
are adjusted to the experimental data
for the fusion or capture cross sections.
To calculate the nucleus-nucleus interaction potential $V(R)$,
we use the procedure presented in Refs.~\cite{EPJSub,EPJSub1}.
For the nuclear part of the nucleus-nucleus
potential, the double-folding formalism with
the Skyrme-type density-dependent effective
nucleon-nucleon interaction is used.

To analyze the experimental date on fusion  cross section, it is useful to use the so called
universal fusion function (UFF) $F_0$ ~\cite{GomesUFF}. The advantages of  UFF appear clearly
when one wants to compare fusion  cross sections for systems with quite different
Coulomb barrier heights and positions. In the reactions where the capture and fusion cross sections
coincide, the comparison of experimental cross
sections with the UFF allows us to make conclusions about the role of deformation of
colliding nuclei and the nucleon transfer between interacting nuclei in the capture cross
section because the UFF (the consequence of the Wong's formula) does not contain these effects.
In Ref.~\cite{GomesUFF} a reduction procedure was proposed to eliminate the influence
of the nucleus-nucleus
potential on the fusion cross section.
It consists of the following transformations:
$$E_{\rm c.m.} \rightarrow x= \dfrac{E_{\rm c.m.}-V_b}{\hbar \omega},
\qquad \sigma^{exp} \rightarrow F(x)=\dfrac{2 E_{\rm c.m.}}{\hbar \omega R_b^{2}}\sigma^{exp}.$$
The frequency $\omega =\sqrt{V^{''}(R_b)/\mu}$
is related with the second derivative $V^{''}(R_b)$ of the total nucleus-nucleus potential $V(R)$
(the Coulomb + nuclear parts)
at the barrier radius $R_b$ and the reduced mass parameter $\mu$.
With these replacements one can compare the experimental data for different reactions.
After these transformations,
 the reduced calculated fusion cross section
takes the simple form
$$F_0=\ln [1+\exp(2\pi x)].$$
To take into consideration the deviation of the real potential from the inverted oscillator,
we modify the reduction procedure as follows:
$$E_{\rm c.m.} \rightarrow x= S/(\hbar\pi),$$
$$ \qquad \sigma^{exp} \rightarrow F(x)= \dfrac{2SE_{\rm c.m.}}{\hbar\pi R_b^2(V_b-E_{\rm c.m.})}\sigma^{exp}.$$
In this case
$$F_0=\ln [1+\exp(-2S/\hbar)],$$
where
$S(E_{\rm c.m.})$ is  the classical action.
At energies above the Coulomb barrier,
we have $S=\pi (V_b - E_{\rm c.m.})/\omega$.

\section{Results of calculations}
\subsection{Effect of quadrupole deformation}
In Fig.~1 (upper part), one can see the comparisons of dependencies
$F$ and $F_0$ on $S/(\hbar\pi)$ for some reactions
considered in present paper. As  expected, at sub-barrier energies
 the deviation from the UFF is larger in
the case of reactions with strongly deformed target-nuclei and large factor
$Z_1 \times Z_2$ ($^{16}$O,$^{40}$Ar,$^{48}$Ca+ $^{154}$Sm, $^{74}$Ge + $^{74}$Ge).
For the reactions   $^{16}$O,$^{40}$Ar+$^{144}$Sm  with spherical
targets the experimental cross sections
are rather close to the UFF.
\begin{figure}
\vspace*{-0.cm}
\centering
\includegraphics[angle=0, width=0.8\columnwidth]{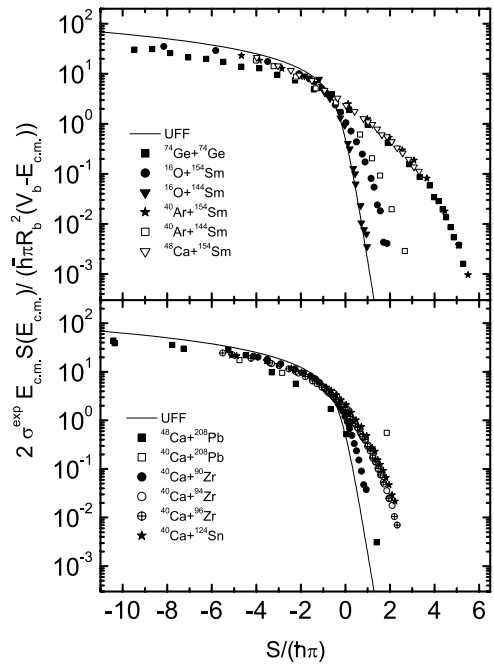}
\vspace*{-0.2cm}
\caption{Comparison of modified UFF  $F_0$ with the experimental values of
$\dfrac{2E_{\rm c.m.}S(E_{\rm c.m.})}{\hbar\pi R_b^2(V_b-E_{\rm c.m.})}\sigma^{exp}$
for the indicated reactions. The experimental data for $\sigma^{exp}$ are  from
Refs.~\protect\cite{KnyazevaCa48Sm154,LeighO16Sm,DiGregO16Sm,BeckermanGe74Ge74,Reisd40ArSmSn,TimmersCa40Zr96,4048Ca208Pb,48CaPb208}.
}
\label{1_fig}
\end{figure}
To separate the effects of deformation and neutron transfer, firstly we consider the
reactions with deformed nuclei in which $Q$-value for the neutron transfer are small, i.e.
the neutron transfers can be disregarded.
In Figs.~2 and 3, the calculated capture cross sections for the reactions
$^{16}$O,$^{48}$Ca,$^{40}$Ar+$^{154}$Sm, and
$^{74}$Ge+$^{74}$Ge  are in a good agreement with the available experimental
data~\cite{KnyazevaCa48Sm154,LeighO16Sm,BeckermanGe74Ge74,Reisd40ArSmSn}
showing that the quadrupole deformations of the interacting nuclei
are the main reasons for the enhancement of the  capture cross section
 at sub-barrier energies.
The  quadrupole
deformation parameters $\beta_2$ are taken from Ref.~\cite{Ram} for the deformed even-even nuclei.
In Ref.~\cite{Ram} the quadropole deformation parameters $\beta_2$ for the first excited $2^+$  states of nuclei
are given. For the nuclei deformed in the ground state, the $\beta_2$ in  $2^+$ state is similar to the  $\beta_2$
in the ground state and we use $\beta_2$ from Ref.~\cite{Ram} in the calculations. For double magic nuclei,
in the ground state we take $\beta_2=0$.
In  Ref.~\cite{GomesRec} the experimentally observed enhancement of sub-barrier fusion for the reactions
$^{16}$O,$^{48}$Ca+$^{154}$Sm,  and $^{74}$Ge+$^{74}$Ge
was explained by the nucleon transfer and
neck formation effects. However, in the present article we demonstrate that  a good agreement
with the experimental data at sub-barrier energies could be reached
taking  only the  quadrupole  deformations of interacting nuclei into consideration.
\begin{figure}
\vspace*{-0.0cm}
\centering
\includegraphics[angle=0, width=0.95\columnwidth]{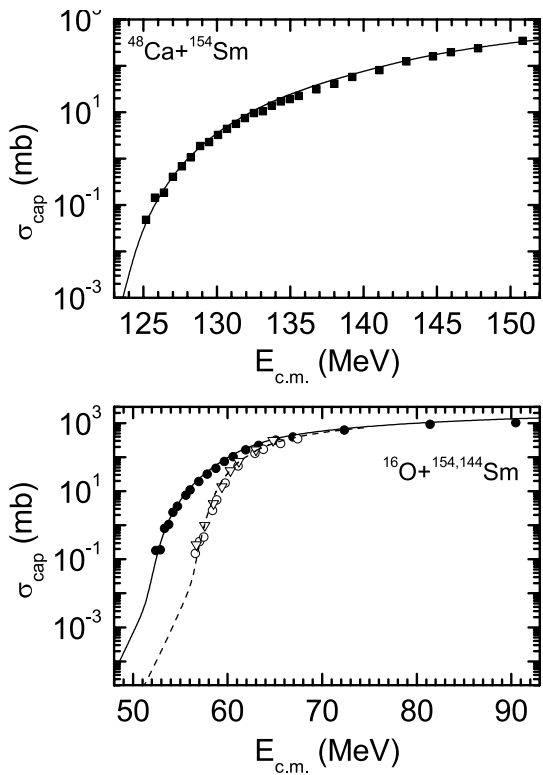}
\vspace*{-0.2cm}
\caption{The calculated capture cross sections versus $E_{\rm c.m.}$ for the indicated reactions
$^{16}$O,$^{48}$Ca + $^{154}$Sm (solid lines), and  $^{16}$O + $^{144}$Sm (dashed line).
The experimental  data (symbols) are  from Refs.~\protect\cite{KnyazevaCa48Sm154,LeighO16Sm,DiGregO16Sm}.
The following  quadrupole deformation parameters  are used:
$\beta_{2}$($^{154}$Sm)=0.341~\protect\cite{Ram},
$\beta_{2}$($^{144}$Sm)=0.05, and $\beta_{2}$($^{16}$O)=$\beta_{2}$($^{48}$Ca)=0.
}
\label{2_fig}
\end{figure}
\begin{figure}
\vspace*{0.cm}
\centering
\includegraphics[angle=0, width=0.95\columnwidth]{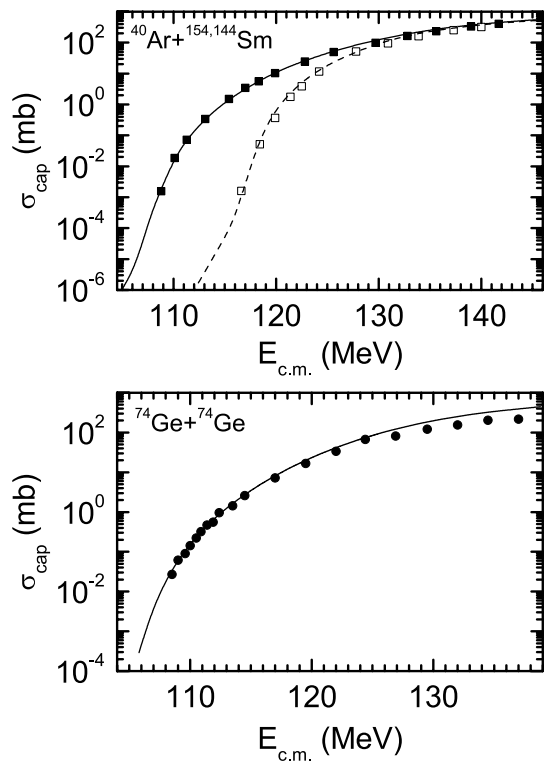}
\vspace*{-0.2cm}
\caption{The same as  Fig.~2,  for the indicated reactions
$^{74}$Ge+$^{74}$Ge, $^{40}$Ar + $^{154}$Sm  (solid lines), and $^{40}$Ar + $^{144}$Sm (dashed line).
The experimental  data  (symbols) are  from Ref.~\protect\cite{BeckermanGe74Ge74,Reisd40ArSmSn}.
The following  quadrupole deformation parameters  are used:
$\beta_{2}$($^{40}$Ar)=0.25~\protect\cite{Ram},
$\beta_{2}$($^{74}$Ge)=0.2825~\protect\cite{Ram},
$\beta_{2}$($^{154}$Sm)=0.341~\protect\cite{Ram},
and $\beta_{2}$($^{144}$Sm)=0.05.
}
\label{3_fig}
\end{figure}
\begin{figure}
\vspace*{0.cm}
\centering
\includegraphics[angle=0, width=0.77\columnwidth]{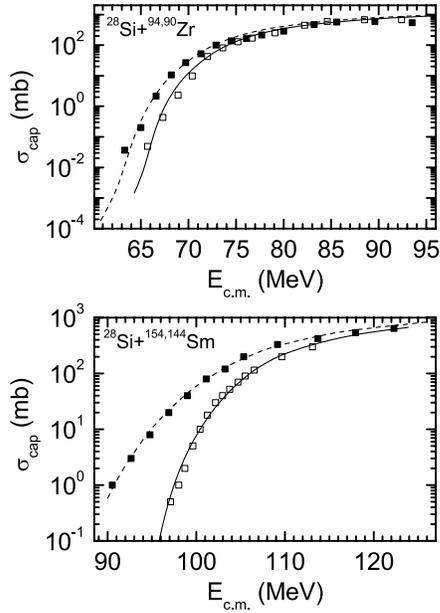}
\vspace*{0.cm}
\caption{The same as  Fig.~2,  for the  indicated  reactions
$^{28}$Si+$^{94}$Zr,$^{154}$Sm  (solid lines),
and $^{28}$Si + $^{90}$Zr,$^{144}$Sm (dashed lines).
The experimental  data   (symbols)  are
from  Refs.~\protect\cite{Kalkal28SiZr9490,GilSi28154sm,NobreSi28144sm}.
The following  quadrupole deformation parameters  are used:
$\beta_{2}$($^{154}$Sm)=0.341~\protect\cite{Ram},
$\beta_{2}$($^{144}$Sm)=0.05, and $\beta_{2}$($^{28}$Si)=0.3.
}
\label{4_fig}
\end{figure}
\begin{figure}
\vspace*{0.0cm}
\centering
\includegraphics[angle=0, width=0.77\columnwidth]{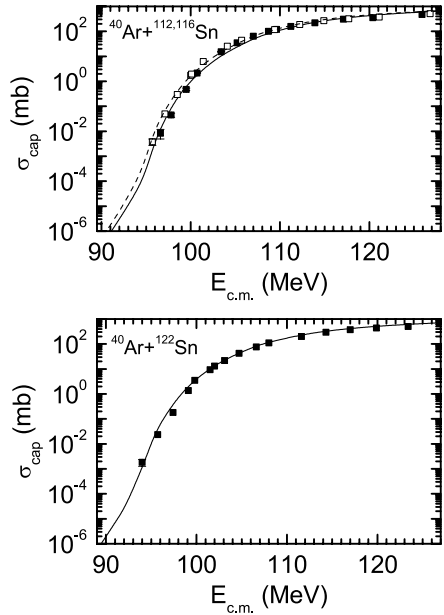}
\vspace*{0.cm}
\caption{The same as  Fig.~2,  for the  indicated  reactions
$^{40}$Ar + $^{112,122}$Sn  (solid lines), and $^{40}$Ar + $^{116}$Sn (dashed line).
The experimental  data (symbols)  are  from
Ref.~\protect\cite{Reisd40ArSmSn}.
The following quadrupole deformation parameters  are used:
$\beta_{2}$($^{112}$Sn)=0.1227~\protect\cite{Ram},
$\beta_{2}$($^{116}$Sn)=0.1118~\protect\cite{Ram},
$\beta_{2}$($^{122}$Sn)=0.1036~\protect\cite{Ram},
and $\beta_{2}$($^{40}$Ar)=0.25~\protect\cite{Ram}.
}
\label{5_fig}
\end{figure}
\begin{figure}
\vspace*{-0.4cm}
\centering
\includegraphics[angle=0, width=0.8\columnwidth]{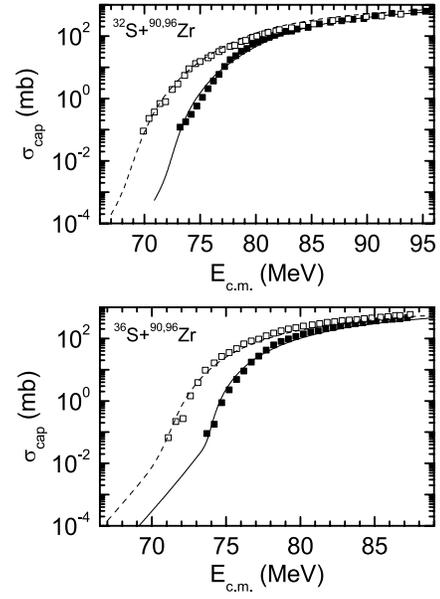}
\vspace*{-0.0cm}
\caption{The same as  Fig.~2,  for the  indicated  reactions
$^{36,32}$S + $^{90}$Zr (solid lines), and $^{36,32}$S + $^{96}$Zr (dashed lines).
The experimental  data  (symbols) are  from
Refs.~\protect\cite{ZhangS32Zn9096,StefaniniS36Zn9096}.
The following  quadrupole deformation parameters  are used:
$\beta_{2}$($^{32}$S)=0.312~\protect\cite{Ram},
$\beta_{2}$($^{34}$S)=0.252~\protect\cite{Ram},
$\beta_{2}$($^{96}$Zr)=0.08, and
$\beta_{2}$($^{36}$S)=$\beta_{2}$($^{90}$Zr)=0.
}
\label{6_fig}
\end{figure}
\begin{figure}
\vspace*{-1.0cm}
\centering
\includegraphics[angle=0, width=0.82\columnwidth]{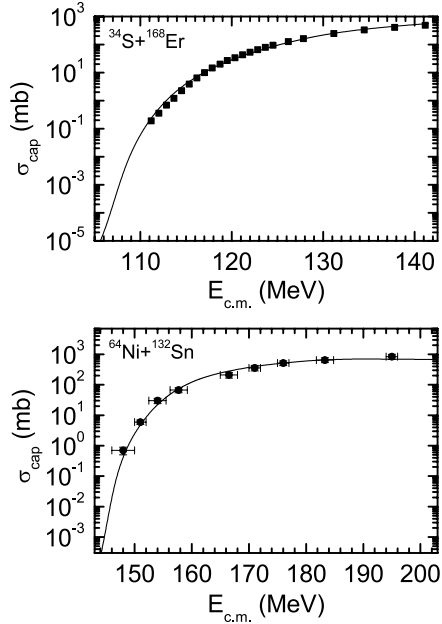}
\vspace*{-0.0cm}
\caption{The same as  Fig.~2,  for the indicated  reactions
$^{34}$S + $^{168}$Er and $^{64}$Ni + $^{132}$Sn.
The experimental  data (symbols) are  from
Refs.~\protect\cite{Morton34S168Er,LiangNi64Sn132}.
The following quadrupole deformation parameters  are used:
$\beta_{2}$($^{168}$Er)=0.3381~\protect\cite{Ram},
$\beta_{2}$($^{66}$Ni)=0.158~\protect\cite{Ram},
$\beta_{2}$($^{130}$Sn)=0,
and
$\beta_{2}$($^{34}$S)=0.125.
}
\label{7_fig}
\end{figure}
\begin{figure}
\vspace*{-0.0cm}
\centering
\includegraphics[angle=0, width=1.0\columnwidth]{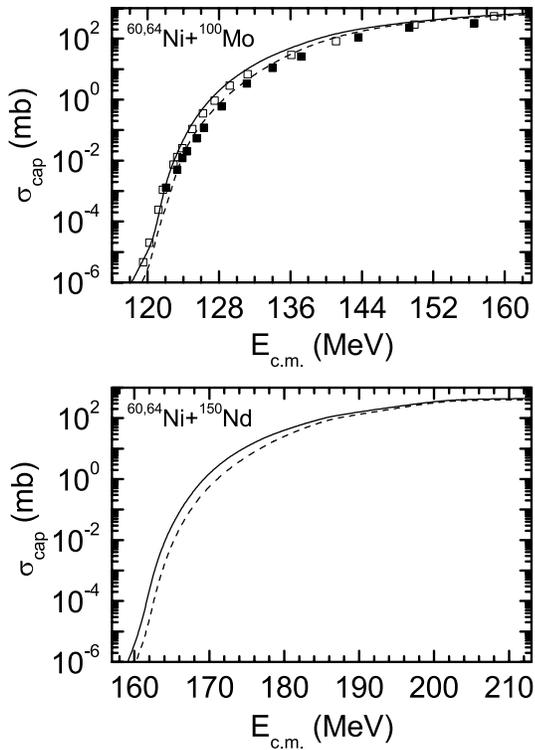}
\vspace*{-0.cm}
\caption{The same as  Fig.~2,  for the  indicated  reactions $^{64}$Ni + $^{100}$Mo,$^{150}$Nd (solid lines)
and $^{60}$Ni + $^{100}$Mo,$^{150}$Nd (dashed lines).
The experimental  data (symbols) for the $^{64}$Ni + $^{100}$Mo reaction
are  from
Ref.~\protect\cite{Jiang64Ni100Mo}.
The following  quadrupole deformation parameters  are used:
$\beta_{2}$($^{62}$Ni)=0.1978~\protect\cite{Ram},
$\beta_{2}$($^{98}$Mo)=0.1684~\protect\cite{Ram},
$\beta_{2}$($^{100}$Mo)=0.2309~\protect\cite{Ram},
$\beta_{2}$($^{148}$Nd)=0.2036~\protect\cite{Ram},
$\beta_{2}$($^{150}$Nd)=0.2848~\protect\cite{Ram},
and $\beta_{2}$($^{64}$Ni)=0.087.
}
\label{8_fig}
\end{figure}
\begin{figure}
\vspace*{-0.cm}
\centering
\includegraphics[angle=0, width=1.0\columnwidth]{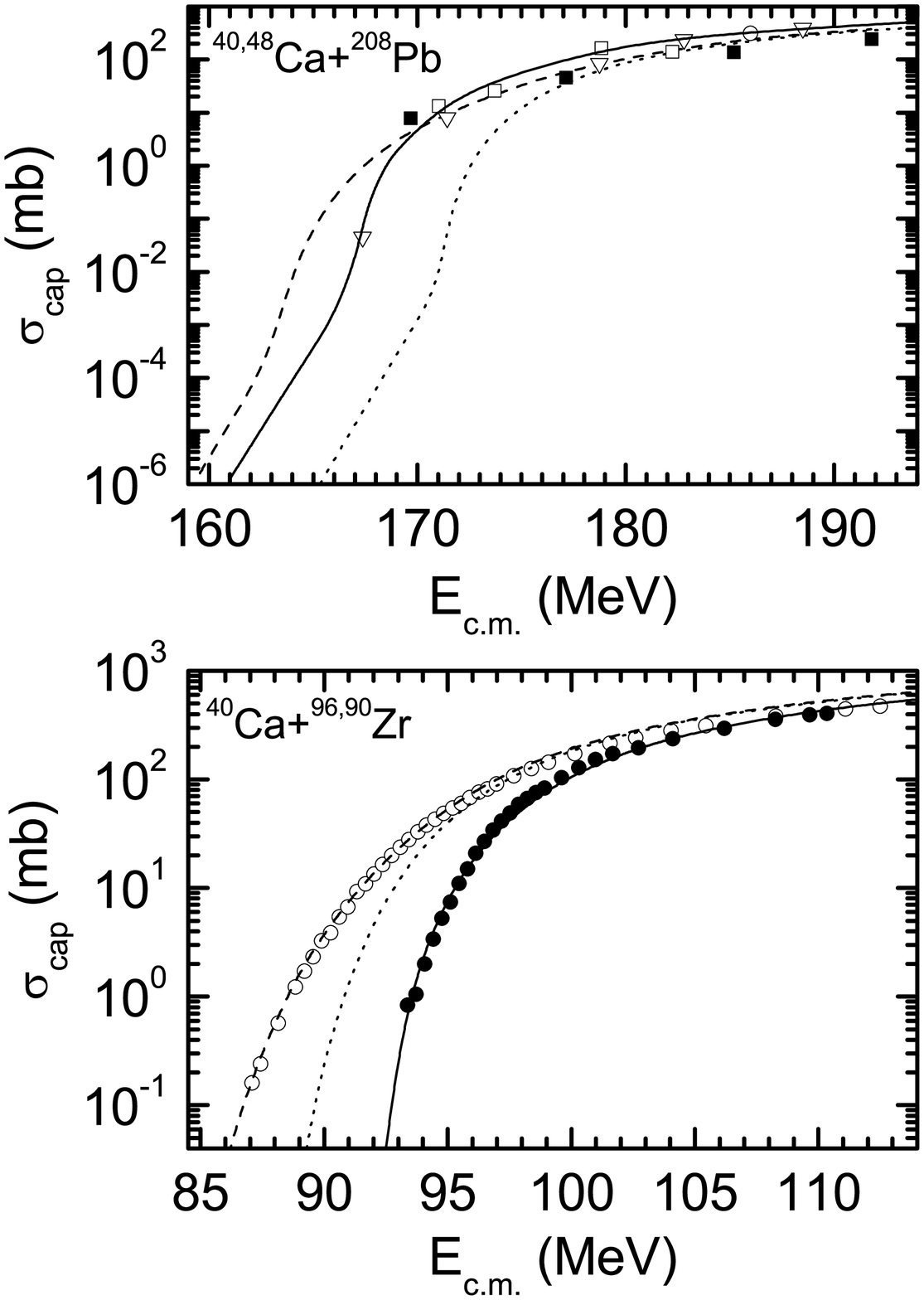}
\vspace*{-0.cm}
\caption{The same as  Fig.~2, for the  indicated  reactions
$^{40}$Ca + $^{96}$Zr,$^{208}$Pb (dashed lines), $^{40}$Ca + $^{90}$Zr (solid line),
and $^{48}$Ca + $^{208}$Pb (solid line and open squares and  triangles).
For the  reactions $^{40}$Ca + $^{96}$Zr,$^{208}$Pb,
the calculated capture cross sections without taking into consideration
the neutron transfer process are shown by dotted lines.
The experimental data (symbols)
are from Refs.~\protect\cite{TimmersCa40Zr96,4048Ca208Pb,48CaPb208}.
The following  quadrupole deformation parameters  are used:
$\beta_{2}$($^{42}$Ca)=0.247~\protect\cite{Ram}, $\beta_{2}$($^{94}$Zr)=0.09~\protect\cite{Ram},
$\beta_{2}$($^{96}$Zr)=0.08, and
$\beta_{2}$($^{40}$Ca)=$\beta_{2}$($^{48}$Ca)=$\beta_{2}$($^{90}$Zr)=$\beta_{2}$($^{206,208}$Pb)=0.
}
\label{9_fig}
\end{figure}
We should mention, that for the sub-barrier energies the results of calculations
are very sensitive to the  quadrupole   deformation parameters $\beta_2$ of the interacting nuclei.
Since there are  uncertainties in the definition of the values of $\beta_2$
in the light- and the medium-mass nuclei,
one can extract the
quadrupole deformation parameters of
these  nuclei from the comparison
of the calculated capture cross sections with the experimental data.
The best case is
when the projectile or target is the spherical double magic nucleus and
there are no neutron transfer channels with large positive $Q$-values.
In this way by describing the reactions $^{28}$Si + $^{90}$Zr,$^{144}$Sm,
$^{34}$S + $^{168}$Er,
$^{36}$S + $^{90,96}$Zr,
$^{40}$Ar + $^{112,116,122}$Sn,$^{144}$Sm,
$^{58}$Ni + $^{58}$Ni, $^{64}$Ni + $^{100}$Mo,$^{74}$Ge  (Figs.~5--10),
we extract the following values of the quadrupole  deformation parameter
$\beta_2$=0.30, 0.125, 0, 0.25, 0.05, 0.087, 0, 0.08, 0.12, 0.11, 0.1, and
0.05   for the  nuclei
$^{28}$Si,   $^{34}$S, $^{36}$S, $^{40}$Ar, $^{58}$Ni, $^{64}$Ni,
$^{90}$Zr, $^{96}$Zr, $^{112}$Sn, $^{116}$Sn, $^{122}$Sn, and  $^{144}$Sm,
respectively.
Note that almost the same values of  quadrupole  deformations
parameters of nuclei in  the ground state were predicted within the mean-field and the macroscopic-microscopic
models~\cite{Pet}.
For  $^{40}$Ar,  $^{96}$Zr, $^{112}$Sn, $^{116}$Sn, and $^{122}$Sn
the extracted $\beta_2$ for  are equal to the experimental ones
from Ref.~\cite{Ram}. These extracted deformation parameters we use in calculations in
next subsection.
Note that almost the same values of  quadrupole  deformations
parameters of nuclei in  the ground state
were predicted within the mean-field and the macroscopic-microscopic
models~\cite{Pet}.
For  $^{40}$Ar,  $^{96}$Zr, $^{112}$Sn, $^{116}$Sn, and $^{122}$Sn
the extracted $\beta_2$ for  are equal to the experimental ones
from Ref.~\cite{Ram}.
These extracted deformation parameters we use in calculations in
next subsection.

\subsection{Effect of neutron transfer}
Several experiments were performed to understand the effect
of neutron transfer in the fusion (capture) reactions.
\begin{figure}
\vspace*{0.15cm}
\centering
\includegraphics[angle=0, width=1.0\columnwidth]{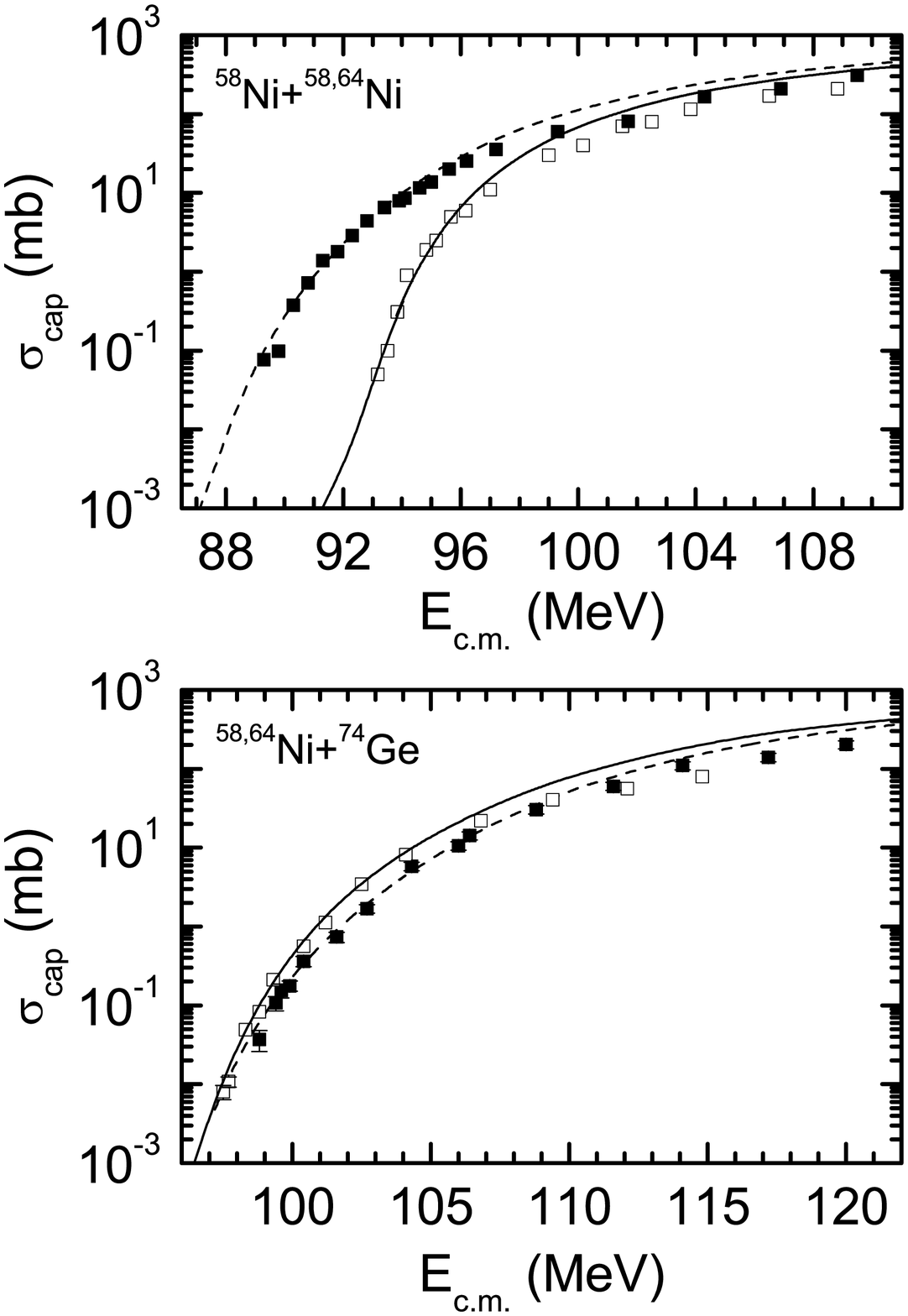}
\vspace*{-0.0cm}
\caption{The same as  Fig.~2, for the indicated   reactions
$^{58}$Ni + $^{64}$Ni,$^{74}$Ge (dashed lines)
and $^{58}$Ni + $^{58}$Ni, $^{64}$Ni + $^{74}$Ge (solid lines).
The experimental data (symbols)
are from Ref.~\protect\cite{Beckerman58Ni5864Ni74Ge}.
The following  quadrupole deformation parameters  are used:
$\beta_{2}$($^{60}$Ni)=0.207~\protect\cite{Ram},
$\beta_{2}$($^{72}$Ge)=0.2424~\protect\cite{Ram},
$\beta_{2}$($^{74}$Ge)=0.2825~\protect\cite{Ram},
$\beta_{2}$($^{58}$Ni)=0.05, and
$\beta_{2}$($^{62}$Ni)$\approx$
$\beta_{2}$($^{64}$Ni)=0.087.
}
\label{10_fig}
\end{figure}
\begin{figure}
\vspace*{0.2cm}
\centering
\includegraphics[angle=0, width=1.0\columnwidth]{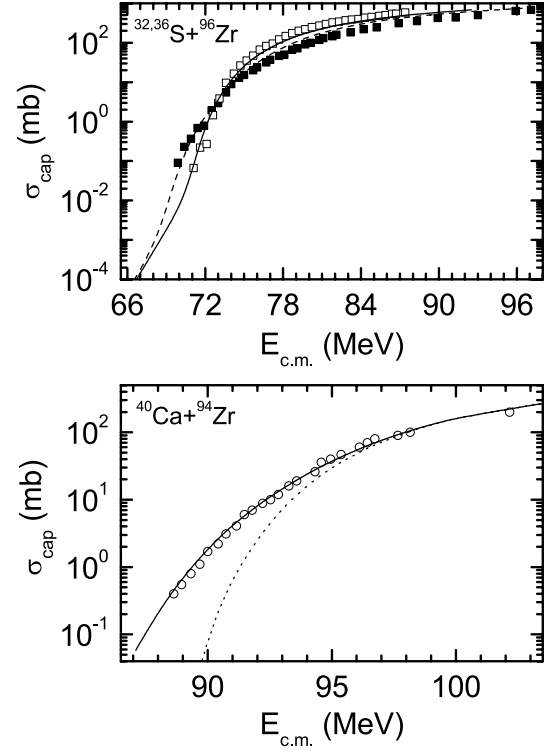}
\vspace*{0.cm}
\caption{The same as  Fig.~2, for the  indicated  reactions
$^{40}$Ca + $^{94}$Zr (solid line),  $^{32}$S + $^{96}$Zr (dashed line and solid squares),
and $^{36}$S + $^{96}$Zr (solid line and open squares).
For the  $^{40}$Ca + $^{94}$Zr reaction,
the calculated capture cross sections without taking into consideration
the neutron transfer process are shown by dotted line.
The experimental data (symbols)
are from Refs.~\protect\cite{ZhangS32Zn9096,StefaniniS36Zn9096,StefaniniCa40Zn94}.
The following  quadrupole deformation parameters  are used:
$\beta_{2}$($^{42}$Ca)=0.247~\protect\cite{Ram},
$\beta_{2}$($^{94}$Zr)=0.09~\protect\cite{Ram},
$\beta_{2}$($^{92}$Zr)=0.1028~\protect\cite{Ram},
$\beta_{2}$($^{96}$Zr)=0.08, and
$\beta_{2}$($^{36}$S)=$\beta_{2}$($^{40}$Ca)=0.
}
\label{11_fig}
\end{figure}
\begin{figure}
\vspace*{-0.4cm}
\centering
\includegraphics[angle=0, width=0.9\columnwidth]{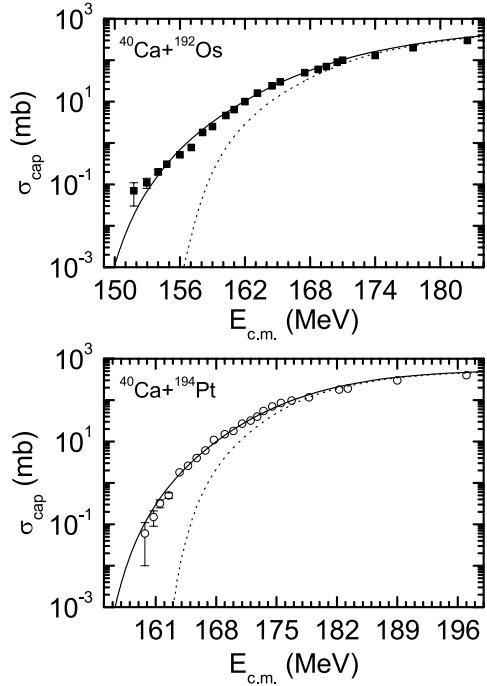}
\vspace*{-0.cm}
\caption{The same as  Fig.~2, for the  indicated  reactions
$^{40}$Ca + $^{192}$Os,$^{194}$Pt (solid lines).
The calculated capture cross sections without taking into consideration
the neutron  transfer process are shown by dotted lines.
The experimental data (symbols)
are from Ref.~\protect\cite{Bierman40ca192Os194Pt}.
The following quadrupole deformation parameters  are used:
$\beta_{2}$($^{42}$Ca)=0.247~\protect\cite{Ram},
$\beta_{2}$($^{192}$Os)=0.1667~\protect\cite{Ram}, $\beta_{2}$($^{190}$Os)=0.1775~\protect\cite{Ram},
$\beta_{2}$($^{194}$Pt)=0.1426~\protect\cite{Ram}, $\beta_{2}$($^{192}$Pt)=0.1532~\protect\cite{Ram},
 and
$\beta_{2}$($^{40}$Ca)=0.
}
\label{12_fig}
\end{figure}
\begin{figure}
\vspace*{-0.0cm}
\centering
\includegraphics[angle=0, width=0.9\columnwidth]{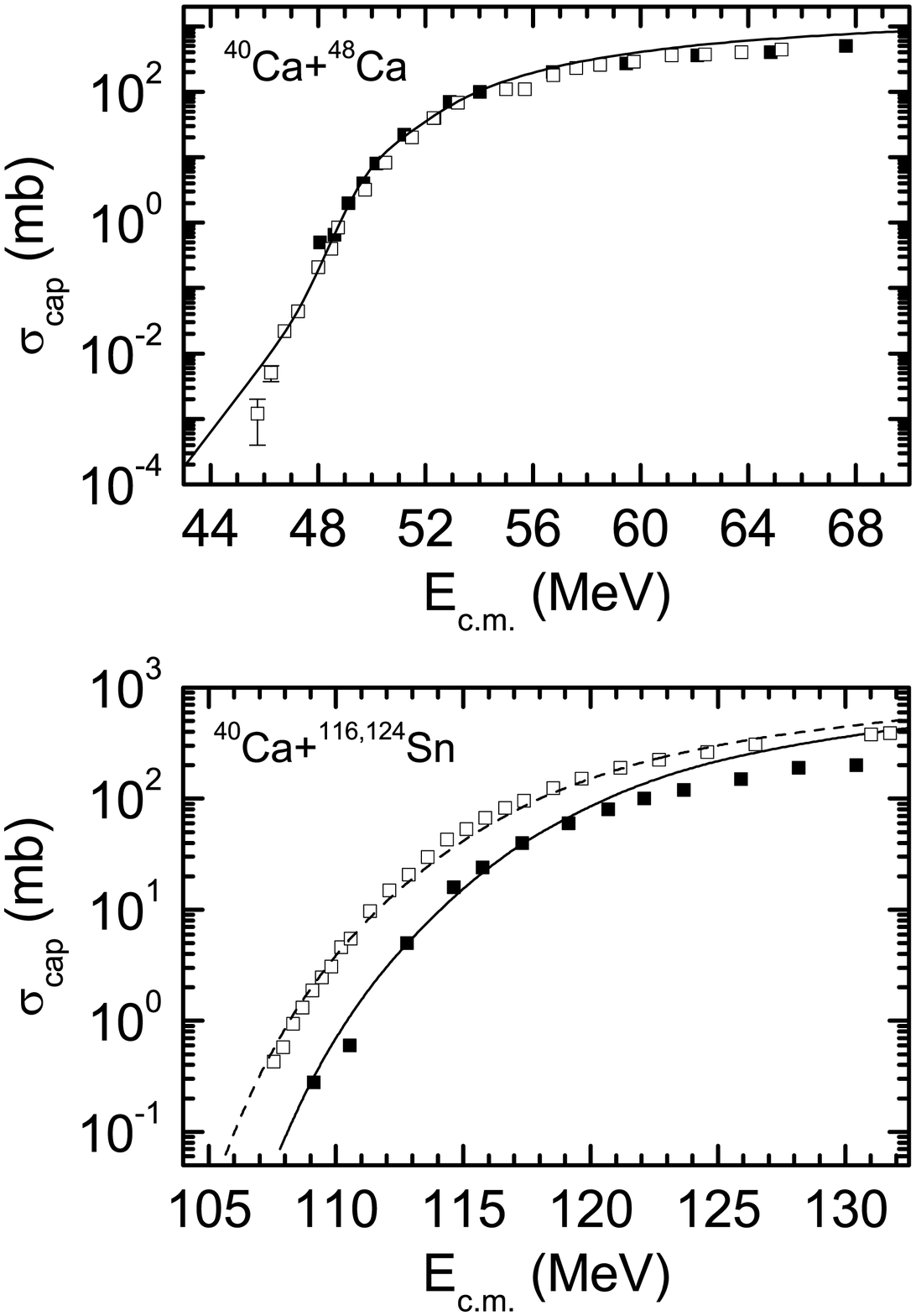}
\vspace*{-0.3cm}
\caption{The same as  Fig.~2, for the  indicated   reactions
$^{40}$Ca + $^{48}$Ca,$^{116}$Sn (solid lines), and $^{40}$Ca + $^{124}$Sn (dashed line).
The experimental data (symbols)
are from Refs.~\protect\cite{Aljuwair40Ca48Ca,trotta40ca48ca,Stefanini40ca116124sn}.
The following  quadrupole deformation parameters  are used:
$\beta_{2}$($^{42}$Ca)=0.247~\protect\cite{Ram},
$\beta_{2}$($^{116}$Sn)=0.1118~\protect\cite{Ram},
$\beta_{2}$($^{122}$Sn)=0.1036~\protect\cite{Ram},
 and $\beta_{2}$($^{46}$Ca)=$\beta_{2}$($^{40}$Ca)=0.
}
\label{13_fig}
\end{figure}
The choice of the projectile-target combination is crucial, and for the systems
studied one can  make unambiguous statements
regarding the neutron transfer process with a positive $Q$-value when the interacting nuclei
are  double magic or semi-magic spherical nuclei.
In this case one can disregard the strong  nuclear deformation effects. The good  examples are
the reactions with the spherical nuclei: $^{40}$Ca + $^{208}$Pb ($Q_{2n}$=5.7 MeV) and
$^{40}$Ca + $^{96}$Zr ($Q_{2n}$=5.5 MeV).
In Fig.~1 (lower part),
one can see that the reduced capture cross sections in these reactions
strongly deviate from the UFF in contrast to those in the reactions
$^{48}$Ca + $^{208}$Pb  and  $^{48}$Ca + $^{96}$Zr,
where the neutron transfer channels are suppressed (the negative $Q$-values).
Since the transfer of protons is shielded by the Coulomb barrier, it occurs when
two nuclei almost touch each other~\cite{obzor}, i.e. after a capture. Thus, the proton transfer can be disregarded
in the calculations of  capture cross sections.
Following the hypothesis of Ref.~\cite{Broglia},
we assume that the sub-barrier capture  mainly  depends  on the two-neutron
transfer with the  positive and relatively large $Q$-value.
Our assumption is that,  before the projectile is captured by target-nucleus
(before the crossing of the Coulomb barrier) which is the slow process,
the two-neutron transfer occurs  at larger separations that can lead to the
population of the first 2$^{+}$ state in the recipient nucleus~\cite{SSzilner}.
\begin{figure}
\vspace*{-1.8cm}
\centering
\includegraphics[angle=0, width=0.85\columnwidth]{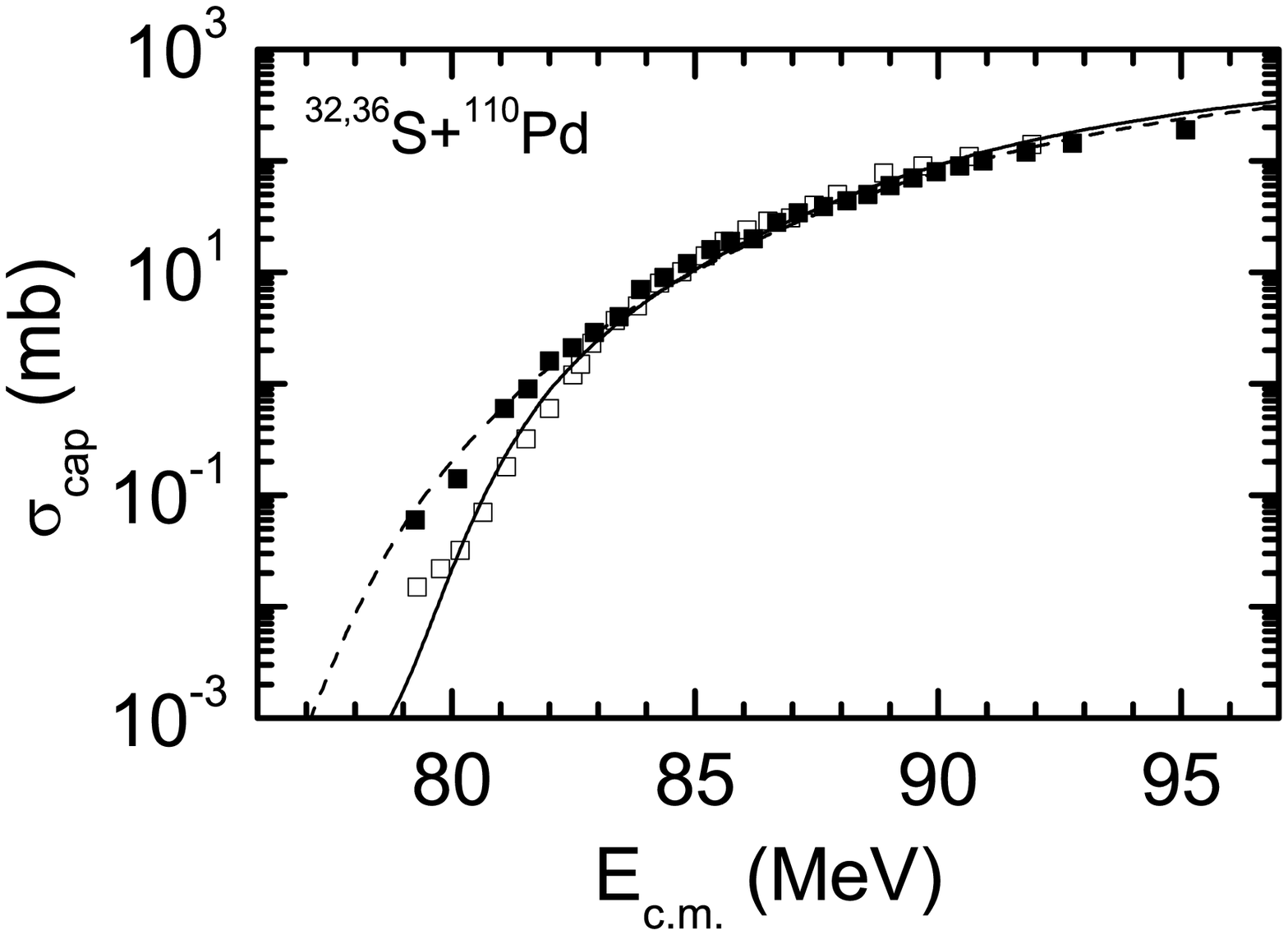}
\vspace*{-0.cm}
\caption{The same as  Fig.~2, for the  indicated  reactions
$^{32}$S + $^{110}$Pd (dashed line and closed squares) and $^{36}$S + $^{110}$Pd  (solid line
and open squares).
The experimental data (symbols)
are from Ref.~\protect\cite{Stefanini3236s110pd}.
The following quadrupole deformation parameters  are used:
$\beta_{2}$($^{34}$S)=0.252~\protect\cite{Ram},
$\beta_{2}$($^{108}$Pd)=0.243~\protect\cite{Ram}, $\beta_{2}$($^{110}$Pd)=0.257~\protect\cite{Ram},
 and
$\beta_{2}$($^{36}$S)=0.
}
\label{14_fig}
\end{figure}
\begin{figure}
\vspace*{-0.cm}
\centering
\includegraphics[angle=0, width=0.77\columnwidth]{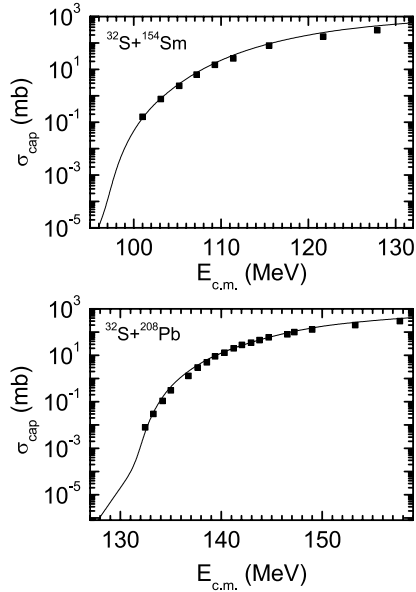}
\vspace*{-0.cm}
\caption{The same as  Fig.~2, for the  indicated  reactions
$^{32}$S + $^{154}$Sm,$^{208}$Pb.
The experimental data (symbols)
are from Refs.~\protect\cite{Stefanini3236s110pd}.
The following  quadrupole deformation parameters  are used:
$\beta_{2}$($^{34}$S)=0.252~\protect\cite{Ram},
$\beta_{2}$($^{152}$Sm)=0.3064~\protect\cite{Ram}, and
$\beta_{2}$($^{206}$Pb)=0.
}
\label{15_fig}
\end{figure}
\begin{figure}
\vspace*{-0.cm}
\centering
\includegraphics[angle=0, width=0.77\columnwidth]{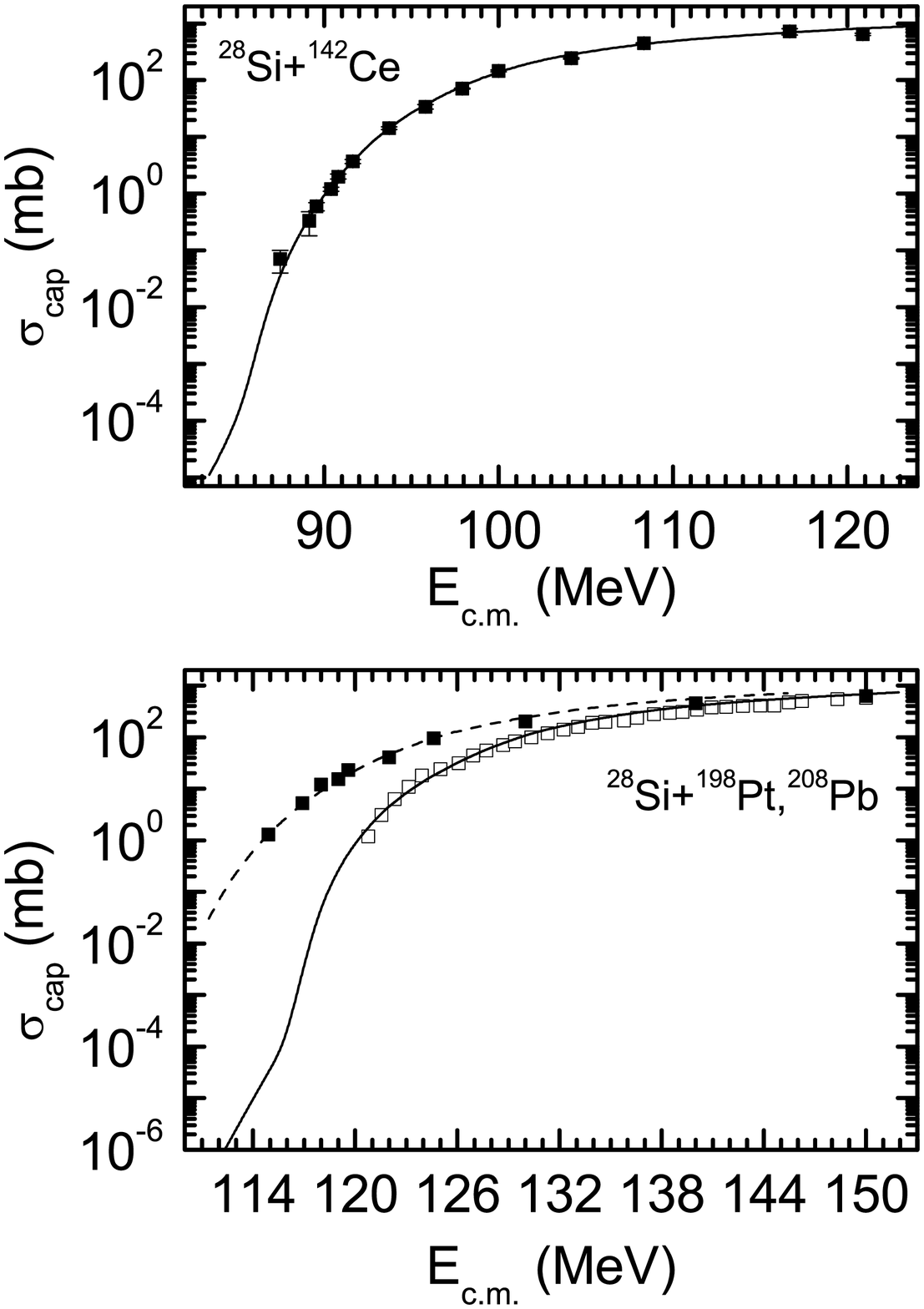}
\vspace*{-0.cm}
\caption{The same as  Fig.~2, for the  indicated  reactions
$^{28}$Si + $^{142}$Ce,$^{208}$Pb (solid lines), and $^{28}$Si + $^{198}$Pt   (dashed line).
The experimental data (symbols)
are from Refs.~\protect\cite{GilSi28Ce142,NishioSi28Pt198,HindeSi28Pb208}.
The following  quadrupole deformation parameters  are used:
$\beta_{2}$($^{30}$Si)=0.315~\protect\cite{Ram},
$\beta_{2}$($^{140}$Ce)=0.1012~\protect\cite{Ram},
$\beta_{2}$($^{196}$Pt)=0.1296~\protect\cite{Ram}, and
$\beta_{2}$($^{206}$Pb)=0.
}
\label{16_fig}
\end{figure}
Since after two-neutron  transfer the mass numbers,  the deformation parameters
of interacting nuclei, and, respectively, the height and shape of the Coulomb barrier are changed,
one can expect the enhancement or suppression of the capture. For example, after the neutron
transfer in the reaction
 $^{40}$Ca($\beta_2=0$) + $^{208}$Pb($\beta_2=0$)$\to ^{42}$Ca($\beta_2=0.247$) + $^{206}$Pb($\beta_2=0$)
($^{40}$Ca($\beta_2=0$) + $^{96}$Zr($\beta_2=0.08$)$\to ^{42}$Ca($\beta_2=0.247$) + $^{94}$Zr($\beta_2=0.09$))
the deformation of the nuclei increases and the mass asymmetry of the system decreases  and
thus the value of the Coulomb barrier decreases and
the capture cross section becomes larger (Fig.~10).
\begin{figure}
\vspace*{0.2cm}
\centering
\includegraphics[angle=0, width=0.77\columnwidth]{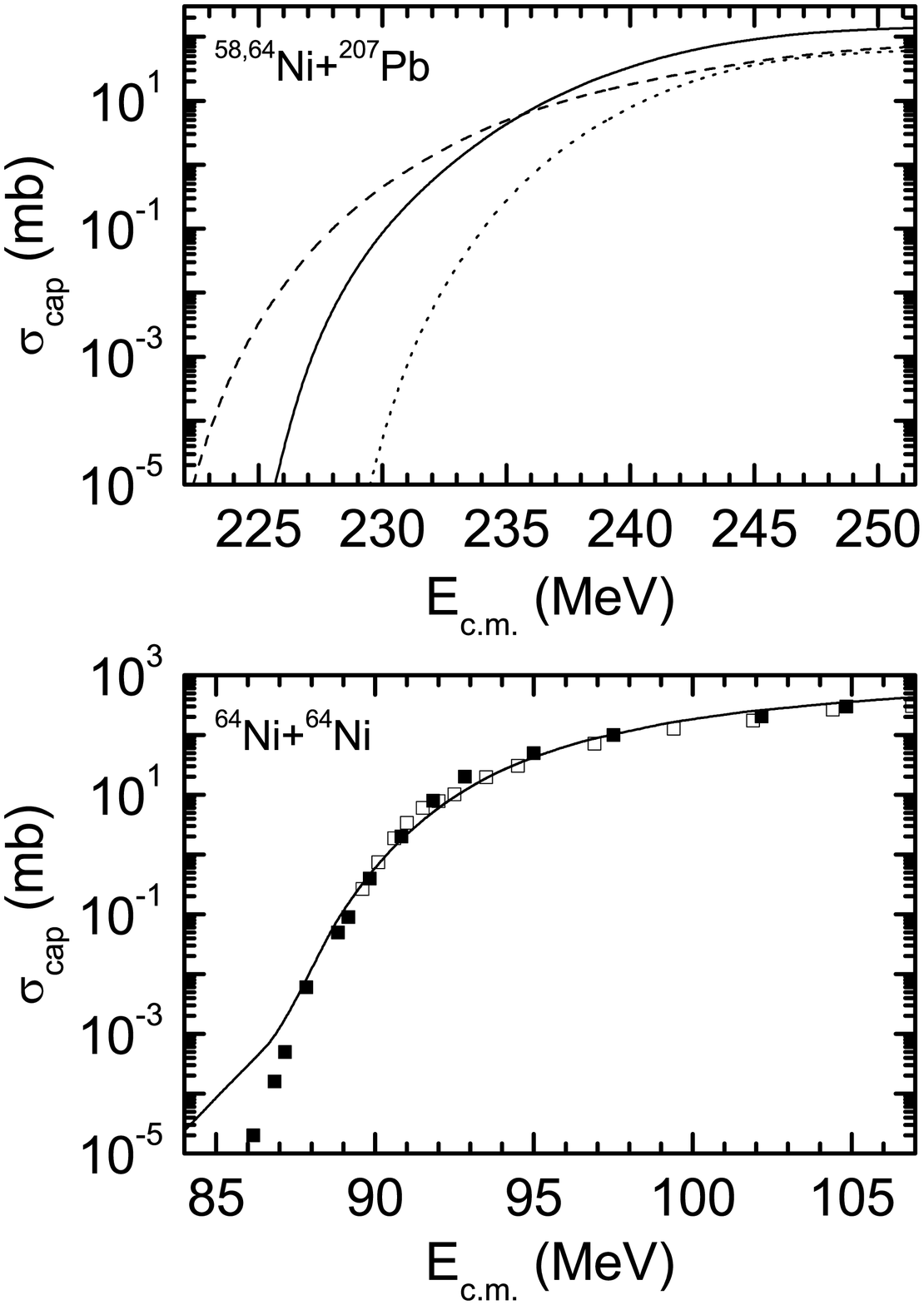}
\vspace*{-0.3cm}
\caption{The same as  Fig.~2, for the  indicated  reactions
$^{58}$Ni + $^{207}$Pb  (dashed line), $^{64}$Ni + $^{64}$Ni (solid line),   and
$^{64}$Ni + $^{207}$Pb  (solid line). For the  $^{58}$Ni + $^{207}$Pb reaction,
the calculated capture cross sections without taking into consideration
the neutron transfer process are shown by dotted line.
The experimental data (symbols)
are from Refs.~\protect\cite{Beckerman58Ni5864Ni74Ge,Jiang64Ni64Ni}.
The following  quadrupole deformation parameters  are used:
$\beta_{2}$($^{60}$Ni)=0.207~\protect\cite{Ram},
$\beta_{2}$($^{58}$Ni)=0.05,
$\beta_{2}$($^{64}$Ni)=0.087,
 and
$\beta_{2}$($^{205,207}$Pb)=0.
}
\label{17_fig}
\end{figure}
We observe the same behavior in the reactions
$^{64}$Ni + $^{132}$Sn (Fig.~7),
$^{58}$Ni+$^{64}$Ni,$^{74}$Ge (Fig.~9), $^{32}$S+$^{96}$Zr, $^{40}$Ca+$^{94}$Zr (Fig.~11),
$^{40}$Ca+$^{192}$Os,$^{198}$Pt (Fig.~12), and $^{40}$Ca + $^{48}$Ca,$^{116,124}$Sn (Fig.~13).
One can see a good agreement between the  calculated results and the experimental data.
For some reactions at energies above the
Coulomb barrier,
the small deviation between the calculated results and experimental data  probably  arises
from the fact that the fusion-fission channel was not taken into consideration
in the experimental capture  cross sections.
So, our results show
that the observed capture enhancement at sub-barrier energies for
the reactions mentioned above
is related to the two-neutron transfer channel.
For these reactions there is a large deflection from the UFF (see lower part of Fig.~1).
Note that strong population of the yrast states, and in particular of the first 2$^+$ state of even Ar (Ca) isotopes
via the neutron pick-up channels in the $^{40}$Ar +  $^{208}$Pb ($^{40}$Ca +  $^{96}$Zr) reaction
 is experimentally found in
Ref.~\cite{SSzilner}.
In the calculations, for such excited recipient nuclei we use the experimental deformation parameters
$\beta_2$ related to the first 2$^+$ states from the table of Ref.~\cite{Ram}. We assume that after two neutron
transfer the residues of donor nuclei remain in the ground state with corresponding quadrupole deformation.

One can find the reactions with large positive two-neutron
transfer $Q$-values where the transfer weakly influences   or even suppresses
the capture process. This happens if after transfer the deformations of nuclei almost do
not  change or even decrease. For instance, in the reactions
%
%
%
%
$^{32}$S($\beta_2=0.312$) + $^{96}$Zr($\beta_2=0.08$)$\to ^{34}$S($\beta_2=0.252$) + $^{94}$Zr($\beta_2=0.09$),
$^{60}$Ni($0.05 < \beta_2 \lesssim 0.1$) + $^{100}$Mo($\beta_2=0.231$)$\to ^{62}$Ni($\beta_2=0.198$) + $^{98}$Mo($\beta_2=0.168$)
and $^{60}$Ni($0.05 < \beta_2 \lesssim 0.1$) + $^{150}$Nd($\beta_2=0.285$)$\to ^{62}$Ni($\beta_2=0.198$) + $^{148}$Nd($\beta_2=0.204$)
one can expect weak dependence of the capture cross section on the neutron transfer~(Figs.~8 and 11).
%
%
%
%
There is the experimental indication of such effect for the $^{60}$Ni + $^{100}$Mo
reaction~\cite{Scarlassara}.
The weak influence of  neutron transfer on the capture process is also found in the
reactions  $^{32}$S + $^{110}$Pd ,$^{154}$Sm,$^{208}$Pb (Figs.~14 and 15),
$^{28}$Si + $^{94}$Zr,$^{142}$Ce,$^{154}$Sm,$^{208}$Pb (Figs.~4 and 16).
The same behaviour is expected in
the reactions $^{84}$Kr +  $^{138}$Ce,$^{140}$Nd.
For these  reactions, the effect of
 quadrupole deformations of  interacting nuclei
is much stronger than the effect of neutron transfer
between the interacting nuclei.

Note that our model predicts almost the same capture cross sections
for the reactions with positive $Q$-values
$^{6}$He,$^{9}$Li,$^{11}$Be +  $^{206}$Pb, $^{18}$O + $^{58}$Ni and for the
reactions without neutron transfer
 $^{4}$He,$^{7}$Li,$^{9}$Be +  $^{208}$Pb,
 $^{16}$O +  $^{60}$Ni, respectively.

In Fig.~17, the  capture cross sections for the reactions
$^{58,64}$Ni + $^{207}$Pb are predicted.
As  seen, there is considerable difference between the capture cross sections in these two reactions
because  of the existence of the two-neutron transfer channel ($Q_{2n}$=5.6 MeV) in the
reaction
$^{58}$Ni + $^{207}$Pb$\to ^{60}$Ni + $^{205}$Pb.
Thus,  the study of these reactions could be a good test for the conclusion about the effect of
neutron transfer. It will be interesting to compare the role of the neutron transfer channel in the
reactions
with spherical nuclei mentioned above  (Fig.~10)
and with deformed targets, $^{40}$Ca + $^{154}$Sm,$^{238}$U (Fig.~18).

Due to a change of the regime of interaction (the turning-off of the nuclear forces and friction)
at sub-barrier energies~\cite{EPJSub,EPJSub1,Conf}, the curve related to the  capture cross section
as a function of bombarding energy has smaller slope~(see Figs.~2--8,10,11,13--16).
This effect is more visible in the capture of  spherical nuclei
without the neutron transfer. However,
the present experimental data at strongly sub-barrier energies are rather poor.

\section{Origin of fusion hindrance in  reactions with medium-mass nuclei at deep  sub-barrier energies}
Since the sum of the fusion cross section $\sigma_{fus}$
and the quasifission cross section $\sigma_{qf}$ gives the capture cross section
$$\sigma_{cap}=\sigma_{fus}+\sigma_{qf},$$
one can estimate the relative contributions of $\sigma_{fus}$ and  $\sigma_{qf}$ to $\sigma_{cap}$.
In Figs.~17, 13 and 19 the calculated capture cross section are presented for
the reactions $^{40}$Ca + $^{48}$Ca,  $^{64}$Ni + $^{64}$Ni and $^{36}$S + $^{48}$Ca,$^{64}$Ni.
\begin{figure}
\vspace*{-0.2cm}
\centering
\includegraphics[angle=0, width=0.85\columnwidth]{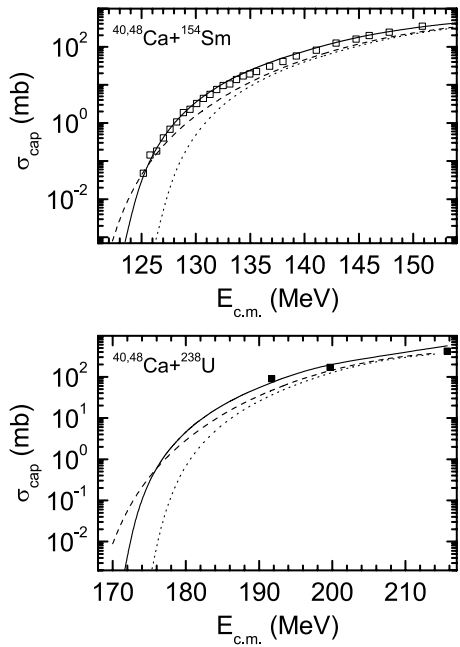}
\vspace*{-0.1cm}
\caption{The same as  Fig.~2, for the  indicated  reactions
$^{40}$Ca + $^{154}$Sm,$^{238}$U  (dashed lines),    and
$^{48}$Ca + $^{154}$Sm,$^{238}$U   (solid lines).
For the   reactions $^{40}$Ca + $^{154}$Sm,$^{238}$U,
the calculated capture cross sections without taking into consideration
the  neutron transfer process are shown by dotted line.
The experimental data   (symbols) for the reactions    $^{48}$Ca + $^{154}$Sm,$^{238}$U
are from Refs.~\protect\cite{KnyazevaCa48Sm154,Shen}.
The following  quadrupole deformation parameters  are used:
$\beta_{2}$($^{42}$Ca)=0.247~\protect\cite{Ram},
$\beta_{2}$($^{152}$Sm)=0.3055~\protect\cite{Ram},
$\beta_{2}$($^{154}$Sm)=0.341~\protect\cite{Ram},
$\beta_{2}$($^{236}$U)=0.2821~\protect\cite{Ram},
$\beta_{2}$($^{238}$U)=0.2863~\protect\cite{Ram},
 and
$\beta_{2}$($^{48}$Ca)=0.
}
\label{18_fig}
\end{figure}
\begin{figure}
\vspace*{-0.4cm}
\centering
\includegraphics[angle=0, width=0.85\columnwidth]{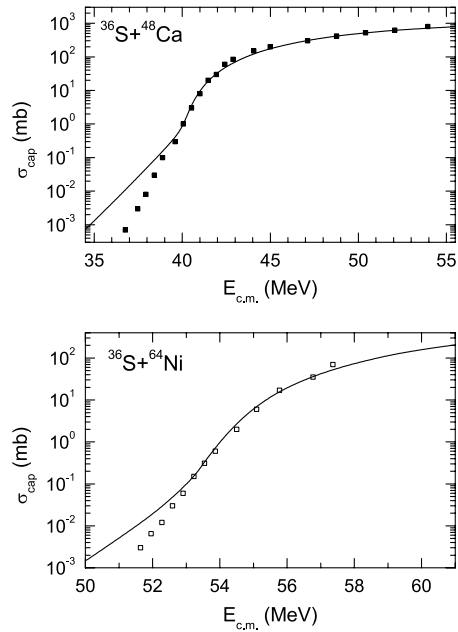}
\vspace*{-0.0cm}
\caption{The same as  Fig.~2, for the  indicated   reactions
$^{36}$S + $^{48}$Ca,$^{64}$Ni.
The experimental data (symbols)
are from Refs.~\protect\cite{StefaniniS36Ca48,MontagnoliS36Ni64}.
The following  quadrupole deformation parameters  are used:
$\beta_{2}$($^{64}$Ni)=0.087 and $\beta_{2}$($^{36}$S)=$\beta_{2}$($^{48}$Ca)=0.
}
\label{19_fig}
\end{figure}
As seen, at energies above and just below the Coulomb barriers
$\sigma_{cap}=\sigma_{fus}$.
The difference between the sub-barrier capture and fusion cross sections
becomes larger with decreasing bombarding energy $E_{\rm c.m.}$.
The same effect one can see for the $^{16}$O + $^{208}$Pb  reaction~\cite{EPJSub}.
Assuming that the estimated capture  and the measured fusion cross sections are correct,
the small fusion cross section at energies well below the Coulomb barrier
may indicate that other reaction channel is preferable and
the system goes to this channel after the capture.
The observed hindrance factor may be understood in term of quasifission
whose cross section should be added to the $\sigma_{fus}$
 to obtain a meaningful comparison with the calculated capture cross section.
At deep sub-barrier energies,
the quasifission event  corresponds to the formation
of a nuclear-molecular state or dinuclear system
with small excitation energy that separates
(in the competition with  the  compound nucleus formation process)
by the quantum tunneling
through the Coulomb barrier
 in a binary event with mass and charge  close
to the entrance channel. In this sense the quasifission is the general phenomenon
which takes place in the reactions with the
massive~\cite{Volkov,nasha,Avaz,GSI}, medium-mass  and,  probably, light nuclei.
%
%
%
For the medium-mass  and light nuclei,
this  reaction channel is expected to be at deep sub-barrier energies
and has to be studied in the future experiments:
from the measurement of the mass (charge) distribution
in the collisions with total momentum transfer
 one can
show the distinct components due to the quasifission.
Because these energies the angular momentum $J<10$,
the angular distribution would have small anisotropy.
The low-energy experimental data would probably provide straight information
since the high-energy data may be shaded by competing nucleon transfer processes.
Note that the binary decay events were already observed experimentally in Ref.~\cite{Wolfs}
for the $^{58}$Ni + $^{124}$Sn reaction at energies below the Coulomb barrier but
assumed to be related to deep-inelastic scattering.
At energies above the Coulomb barrier the hindrance to fusion was revealed in Ref.~\cite{Pollar}
for the reactions $^{58}$Ni + $^{124}$Sn and  $^{16}$O + $^{208}$Pb.
\\

\section{Summary}
The quantum diffusion approach was applied to study
the capture process in the reactions with  deformed and spherical nuclei at sub-barrier energies.
The available experimental data at energies above and below the Coulomb barrier are well described.
As shown,  the experimentally observed sub-barrier fusion enhancement is mainly related  to the quadrupole
deformation of the colliding nuclei and  neutron transfer with large  positive $Q$-value.
The change of the magnitude of the capture cross section after the neutron transfer
occurs due to the change of the deformations of  nuclei.
When after the neutron transfer the deformations of nuclei
do not change or slightly decrease, the neutron transfer weakly influences or even suppresses
the capture process.
It would  be interesting to study such-type  of reactions.

The importance of quasifission near the entrance channel
was noticed for the reactions with medium-mass nuclei
at extreme sub-barrier energies. The quasifission  can explain the difference
between the capture and fusion cross sections. One can try  to check experimentally these predictions.

\section{acknowledgements}
We thank H.~Jia, J.Q. Li, C.J.~Lin, and S.-G.~Zhou for fruitful discussions and  suggestions.
This work was supported by DFG, NSFC, and RFBR.
The IN2P3(France)-JINR(Dubna), MTA(Hungary)-JINR(Dubna) and Polish - JINR(Dubna)
Cooperation Programmes are gratefully acknowledged.\\


\end{document}